\documentclass[a4paper,fleqn,usenatbib]{mnras}

\usepackage{newtxtext,newtxmath}
\usepackage[T1]{fontenc}
\usepackage{ae,aecompl}
\usepackage{graphicx}	% Including figure files
\usepackage{amsmath}	% Advanced maths commands
\usepackage{amssymb}	% Extra maths symbols
\usepackage[dvipsnames]{xcolor}

\title[Expansion and Age of a Nova]{Measuring the Expansion and Age of the Nova Shell IPHASX\,J210204.7+471015}

\author[E.\ Santamar\'\i a et al.]
{E.\ Santamar\'\i a$^{1,2}$\thanks{E-mail: ivan@astro.iam.udg.mx}, 
M.A.\ Guerrero$^{3}$, 
G.\ Ramos-Larios$^{1,2}$, 
L.\ Sabin$^{4}$, 
R.\ V\'azquez$^{4}$,  
\newauthor
M.A.\ G\'omez-Mu\~noz$^{5,6}$,
 and J.A.\,Toal\'{a}$^{7}$
\\
$^1$CUCEI, Universidad de Guadalajara, Blvd. Marcelino Garc\'\i a Barrag\'an 1421, 44430, Guadalajara, Jalisco, Mexico \\
$^2$Instituto de Astronom\'\i a y Meteorolog\'\i a, Dpto.\ de F\'\i sica,
CUCEI, Av.\ Vallarta 2602, 44130, Guadalajara, Jalisco, Mexico\\
$^3$Instituto de Astrof\'\i sica de Andaluc\'\i a, IAA-CSIC, Glorieta de la
Astronom\'\i a s/n, 18008, Granada, Spain\\
$^4$Instituto de Astronom\'\i a, Universidad Nacional Aut\'onoma de M\'exico,
Apdo.\ Postal 877, C.P. 22860, Ensenada, B.C., Mexico \\
$^5$Instituto de Astrof\'\i sica de Canarias (IAC), E-38205, La Laguna, 
Tenerife, Spain \\
$^6$Universidad de La Laguna (ULL), Departamento de Astrof\'\i sica, 
E-38205, La Laguna, Tenerife, Spain\\
$^7$Instituto de Radioastronom\'{i}a y Astrof\'{i}sica (IRyA), UNAM Campus Morelia,
Apartado postal 3-72, 58090 Morelia, Michoac\'{a}n, Mexico
}

\date{Accepted 2018 November 06. Received 2018 November 01; in original form 2018 July 31}

\pubyear{2019}

\begin{document}
\label{firstpage}
\pagerange{\pageref{firstpage}--\pageref{lastpage}}
\maketitle

\begin{abstract}

  The parallax expansion and kinematics of a nova shell can be used to
  assess its age and distance, and to investigate the interaction of
  the ejecta with the circumstellar medium.
  These are key to understand the expansion and dispersal of the nova
  ejecta in the Galaxy.  
Multi-epoch images and high-dispersion spectroscopic observations of the 
recently discovered classical nova shell IPHASX\,J210204.7+471015 around 
a nova-like system have been used to derive a present day expansion rate 
of 0\farcs100~yr$^{-1}$ and an expansion velocity of 285 km~s$^{-1}$.  
These data are combined to obtain a distance of 600 pc to the nova.  
The secular expansion of the nova shell place the event sometime 
between 1850 and 1890, yet it seems to have been missed at that 
time.  
Despite its young age, 130-170 yrs, we found indications that the ejecta has 
already experienced a noticeable deceleration, indicating the interaction of 
this young nova shell with the surrounding medium.
\end{abstract}


%%%%%%%%%%%%%%%%%%%%%%%%%%%%%%%%%%%%%%%%%%%%%%%%%%

%%%%%%%%%%%%%%%%% BODY OF PAPER %%%%%%%%%%%%%%%%%%

\section{Introduction}

Cataclysmic variables (CVs) are binary systems in close orbit in which a
  secondary component, typically a late-type dwarf, giant or sub-giant star,  
transfers H-rich material to the white dwarf (WD)
  main component via an accretion disc
  or directly onto the surface of highly magnetized WDs.
This material builds up onto the WD surface until it reaches
a critical mass limit and experiences a thermonuclear runaway in a
classical nova (CN) event.
Significant amounts of highly processed material ($10^{-5}-10^{-4} M_\odot$)
are ejected at high speeds ($\sim$1000 km~s$^{-1}$).
The expansion of the nebular shell into the interstellar
medium (ISM) can be followed in the next years \citep{Evans_etal1992}.  

Nebular shells around CNe are scarce \citep{Sahman_etal2015}, 
  but their morphologies provide important clues about
  the geometry of the ejecta \citep[e.g., in HR\,Del,][]{HO2003,MD2009} and 
  its interactions with the circumstellar material and newly developed stellar
  winds \citep[e.g., as in DQ\,Her and RR\,Pic,][]{GO1998,VOR2007}.
  They can also be used to determine important properties of the 
  binary system, such as its orbit inclination
  \citep[e.g., in V2491\,Cyg,][]{Retal2011}, whereas the distance
  and age of the nova event can be derived in conjunction with
  kinematic information \citep{GO2000} or via the expansion
  parallax \citep{DD2000}.  
  The shape of the nova shell is probably correlated with the nova
  speed class \citep{SOD1995}.

There are, however, very few studies on the angular expansion of nova
shells.  \citet{Shara_etal2012a} could not detect expansion in
 the nebular remnant of Z\,Cam, while
\citet{Chesneau_etal2012} had to resort to NACO/Very Large Telescope
(VLT) adaptive optics to detect the expansion of the
  nova shell around V1280\,Sco. The nova shell
GK\,Per is probably the best case study, with multiple knots expanding
isotropically at an angular velocity 0\farcs3-0\farcs5~yr$^{-1}$
\citep{Shara_etal2012b,Liimets_etal2012}.  The expansion velocity of
these knots has remained unchanged since their ejection about a
century ago.  This is somehow surprising, because individual knots
reveal notable interactions with each other and with the ISM
\citep{Harvey_etal2016}.

\begin{figure*}
\centering
\includegraphics[width=0.48\textwidth]{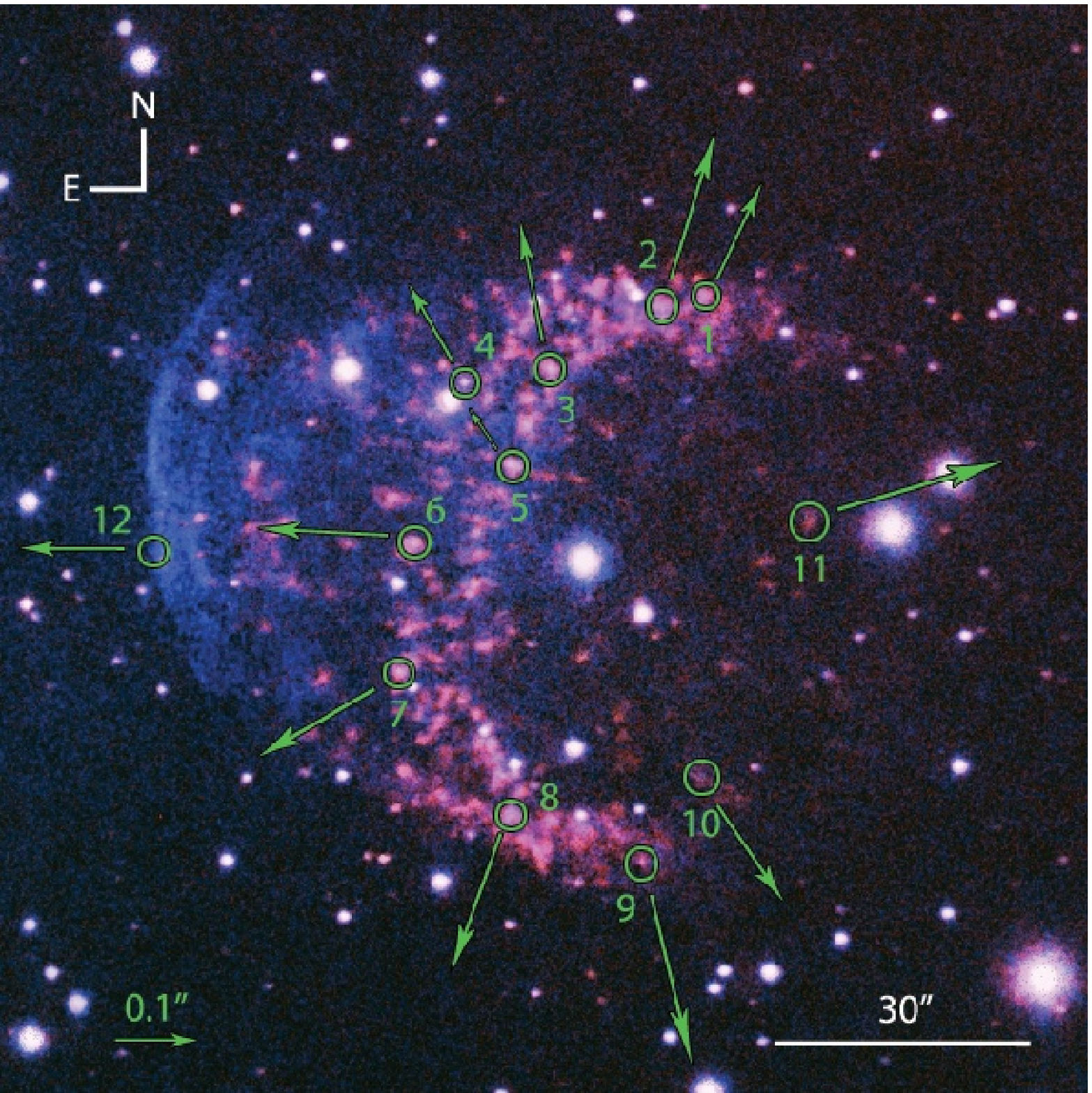}
\hspace{0.1cm}
\includegraphics[width=0.48\textwidth]{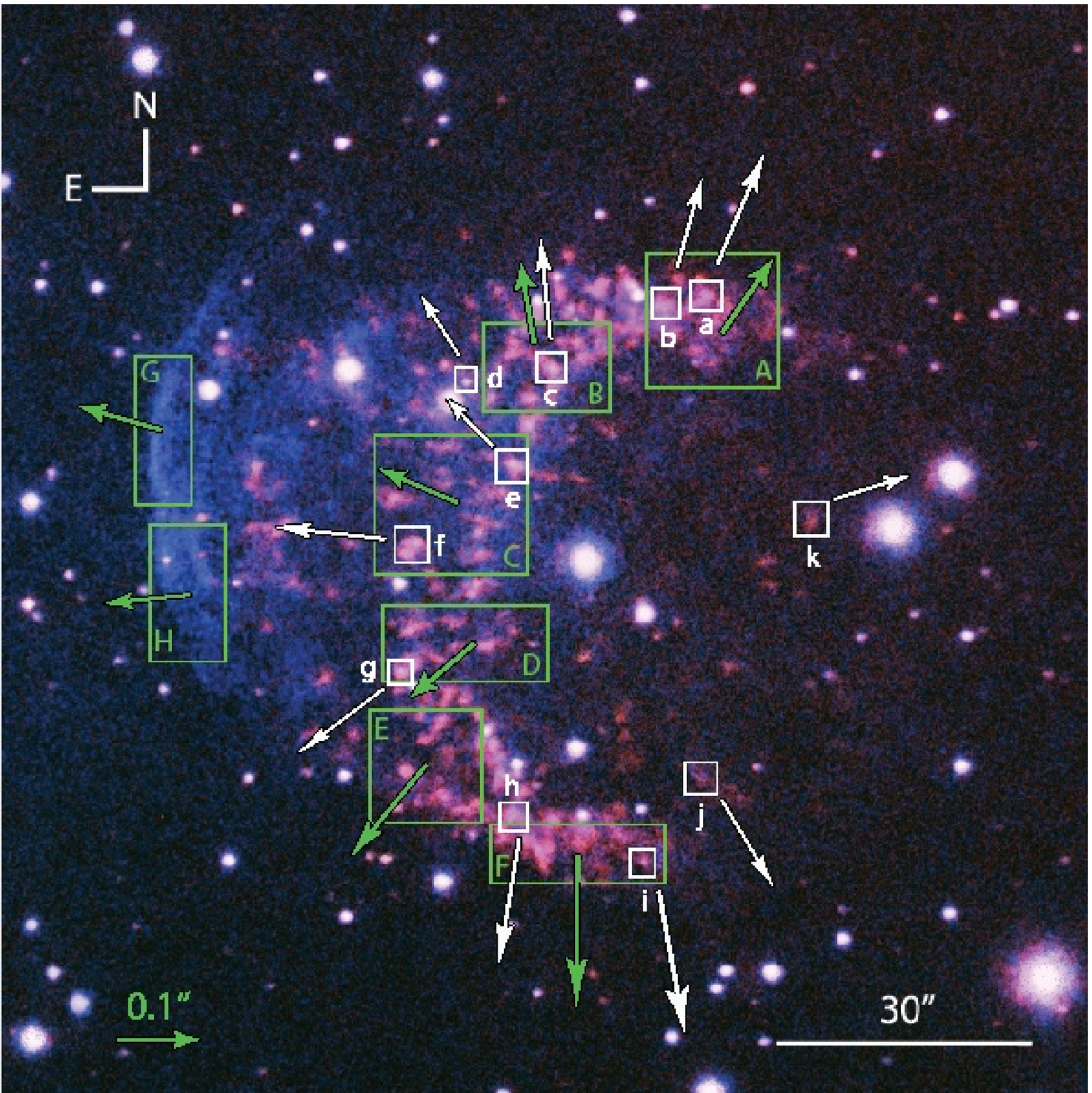}
\caption{
NOT ALFOSC 2017 color-composite picture ([N~{\sc ii}] = red; [O~{\sc iii}] 
= blue) of the nova J210204.
  In this picture, the ring of [N~{\sc ii}]-bright knots shows up in magenta,
  whereas the [O~{\sc iii}]-bright bow-shock does in blue.  
The left and right panel shows the identification of individual features
and boxes used to determine the expansion of the nova using the PCC and QM
methods, respectively (see text for details).  
The size of the arrows in both panels is proportional to the angular
expansion.  
}
\label{fig:img}
\end{figure*}

Since the expansion of a nova shell varies according to the local
properties of the ISM, each object is by itself a case study
offering the possibility to investigate the interaction of highly processed
material with the ISM and its subsequent dispersal into the Galaxy
\citep{Shara_etal2017}.  
Here we present an analysis of the expansion of the recently discovered CN 
shell IPHASX\,J210204.7+471015 \citep[hereafter J210204,][]{Guerrero_etal2018} 
using multi-epoch images and high-dispersion spectra.
The observational material in this work is presented in \S2.
The present day expansion of the nova, its expansion velocity and
distance, and its long-term expansion and age are reported in \S3,
\S4, and \S5, respectively. 
Finally a short discussion of these results are presented in \S6.

\section{Observations} \label{sec:observations}

\subsection{Multi-epoch Imaging}

Contemporary narrow-band images of J210204 were obtained using the 
Andalucia Faint Object Spectrograph and Camera (ALFOSC) attached to the 2.5m 
Nordic Optical Telescope (NOT) of the Roque de los Muchachos Observatory (ORM) 
in La Palma, Spain. 
The images were obtained during three different runs in 2015 July 19,
2016 November 26, and 2017 May 28.  
The E2V 42-40 2k$\times$2k CCD with pixel size 13.5 $\mu$m was used
in 2015, providing a plate scale of 0\farcs189 pix$^{-1}$ and a field
of view (FoV) of 6\farcm5 arcmin, whereas the E2V 231-42 2k$\times$2k
CCD with pixel size 15.0 $\mu$m was used in 2016 and 2017, providing
a plate scale of 0\farcs211 pix$^{-1}$ and a FoV of 7\farcm2 arcmin.  
The 2015 images were acquired through [N~{\sc ii}]
($\lambda_c$=6583 {\AA}, $\Delta\lambda$=36 \AA) and [O~{\sc iii}]
($\lambda_c$=5007 {\AA}, $\Delta\lambda$=43 \AA)
narrow-band filters with total exposure times of 900 and 600 s,
respectively.
The 2016 images were acquired through the same [N~{\sc ii}]
narrow-band filter with total exposure time of 1,800 s.
Although the bandwidth of this [N~{\sc ii}] filter includes the
H$\alpha$ line, the contribution of this line to the [N~{\sc ii}] image
presented here is negligible as the H$\alpha$ line emission is considerably 
weaker than the [N~{\sc ii}] emission \citep{Guerrero_etal2018}.  
The 2017 images were acquired through [N~{\sc ii}]
($\lambda_c$=6584 \AA, $\Delta\lambda$=10 \AA) and [O~{\sc iii}]
($\lambda_c$=5007 \AA, $\Delta\lambda$=30 \AA) narrow-band
filters with total exposure times of 1,800 and 3,600 s, respectively.  
All images were processed using standard {\sc IRAF}\footnote{
  {\sc IRAF} is distributed by the National Optical Astronomy Observatory,
  which is operated by the Association of Universities for Research in
  Astronomy (AURA) under a cooperative agreement with the National Science
  Foundation.
} routines.
  In all cases, images were bias subtracted and flat-field corrected
  using appropriate sky flat field frames.
  Multiple exposures (typically three) were obtained and combined to
  reject cosmic ray hits.
  A small (few arcsec) dither was applied between individual exposures
  to improve the image cosmetics.
The spatial resolution, as determined from the FWHM of field 
stars, was 0\farcs6 in 2015, 0\farcs6 in 2016, and 0\farcs7 
in 2017.

Intermediate-epoch narrow-band H$\alpha$ images of J210204 obtained 
in 2003 and 2009 were downloaded from the IPHAS database
\citep[the INT Photometric H$\alpha$ Survey,][]{Drew_etal2005}. 
The images were obtained using the Wide Field Camera (WFC),
with a plate scale of 0\farcs3 pix$^{-1}$ and FoV 30$^\prime$.  
Exposure times were 120 s, with typical spatial resolution of 
1\farcs4.
An additional 1,200 s INT WFC H$\alpha$ image obtained in 2007 
with similar image quality was also used.

Early broad-band images of J210204 are available in 
the Digitized Sky Survey (DSS).  
A POSS-I-E image was obtained on 1952 July 20 with the Palomar Schmidt 
telescope on a red-sensitive 103aE plate with a plexi filter.
The exposure time was 5 min.
The image has a plate scale of 1\farcs7 pix$^{-1}$ and a spatial
resolution of 3\farcs5.  
A POSS-II-F UKSTU Red image was obtained on 1990 September 19, with the 
Oschin Schmidt Telescope on a IIIaF plate with a RG610 filter.
The exposure time was 6 min. 
The image has a plate scale of 1\farcs0 pix$^{-1}$ and a spatial
resolution of 3\farcs0.  
We emphasize that the DSS broadband images and INT narrowband H$\alpha$
images are mostly dominated by the [N~{\sc ii}] $\lambda\lambda$6548,6584
emission lines, thus allowing a direct comparison with the NOT narrow-band
[N~{\sc ii}] images.  

\subsection{Spectroscopy}

High-dispersion long-slit echelle spectra of J210204 were obtained using
the Manchester Echelle Spectrometer (MES) at the 2.1m telescope of the
Observatorio Astron\'omico Nacional at San Pedro M\'artir, Mexico (OAN-SPM)
on 2016 April 19.  
MES was used with a 6\farcm5-long slit using an H$\alpha$ filter to isolate
the H$\alpha$ and [N~{\sc ii}] $\lambda\lambda$6548,6584 emission lines.  
The E2V~42-40 CCD was used with a 4$\times$4 binning, resulting in
a plate scale of 0\farcs702 pix$^{-1}$ and a spectral scale of 0.1
\AA~pix$^{-1}$.
The slit width of 150 $\mu$m (1\farcs9) implies a spectral
resolution of $\simeq$12$\pm$1 km~s$^{-1}$.
  A single exposure of 1800 s was obtained.
  The data were reduced using standard {\sc IRAF} routines to subtract
  the bias and rectify the spectrum and apply a wavelength calibration
  using a ThAr arc spectrum obtained at the same telescope position as
  the science exposure.  
  The wavelength-calibration resulted in an accuracy $\sim$1 km~s$^{-1}$.

In this work we also use the GTC OSIRIS intermediate-dispersion 
spectra presented by \citet{Guerrero_etal2018}.
Although these data have low spectral dispersion, $\simeq$9 \AA, 
i.e.\ $\approx$400 km~s$^{-1}$ at the [N~{\sc ii}] $\lambda$6584
\AA\ emission line, a Gaussian fit can provide a measurement of
the centroid of a well-detected line with an accuracy $\sim$10\%
the line width.

\section{Current Nova Expansion} \label{sec:today_expansion}

  A color-composite picture of the 2017 images of J210204 is
  shown in Figure~\ref{fig:img}.
  The nova shell J210204 presents a multitude of [N~{\sc ii}]-bright knots
  distributed along a broken ring.
  The morphology of this arc is very similar to the $\sim$3$^\prime$
  in size ring of the shell surrounding the dwarf nova AT\,Cnc
  \citep{Shara_etal2012c}.
  In contrast to that nova shell, J210204 displays an additional
  bow-shock-like structure very prominent in the [O~{\sc iii}]
  emission line.  
  The chemical abundances differences between the ring and bow-shock
  structures imply that the ring is composed of freshly nova ejecta,
  whereas the bow-shock consists mainly of piled up ISM material.  
  The reader is referred to \citet{Guerrero_etal2018} for further details.

\begin{figure}
\centering
\includegraphics[bb=20 90 700 850,width=0.5\textwidth]{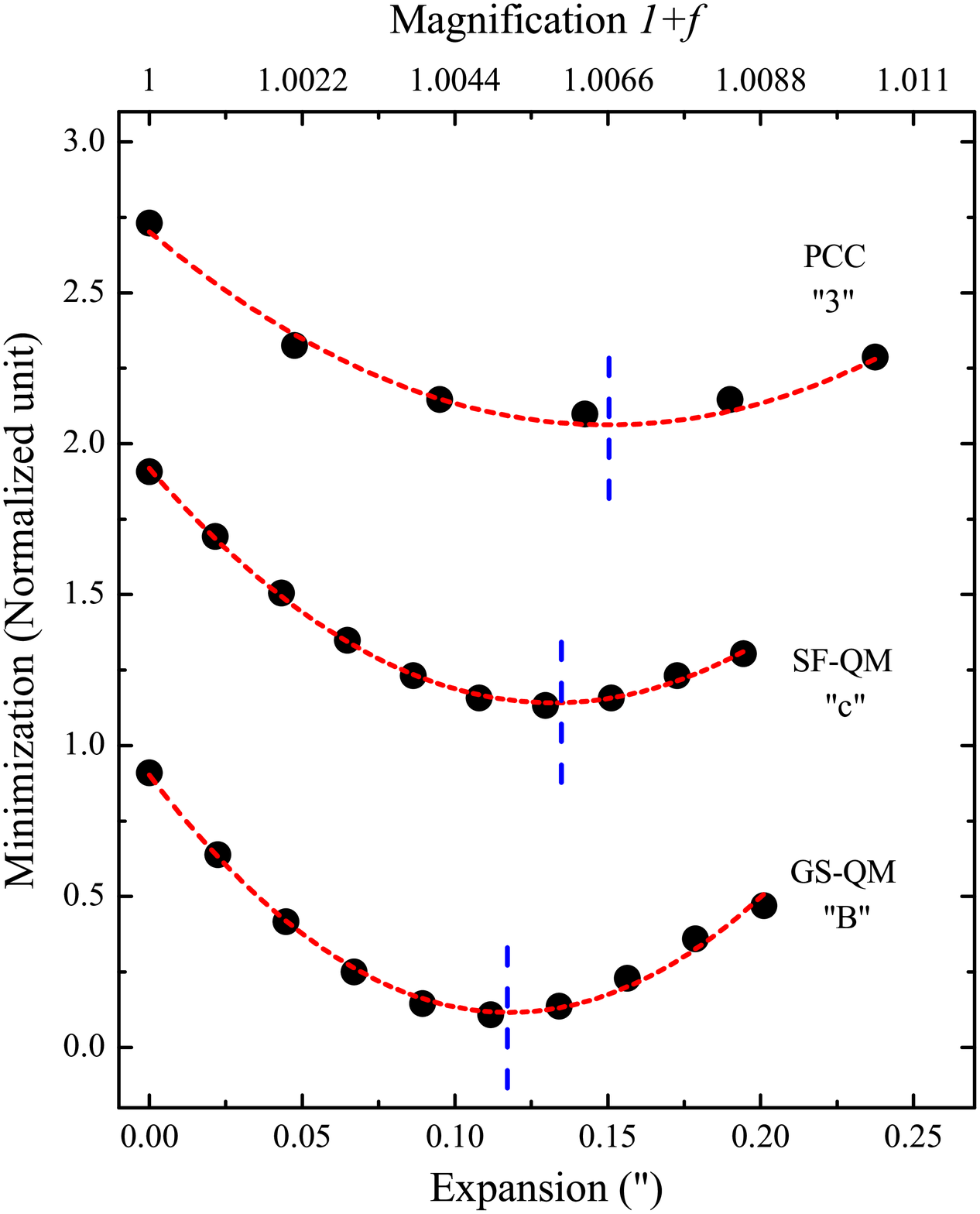}
\caption{
Minimization of the expansion rate and magnification factor for one 
same knot of J210204 using the PCC, SF-QM, and GS-QM methods described 
in the text.  
The knot has label \emph{``3''} in Figure~\ref{fig:img}-{\it left} 
for the PCC method, and labels 
\emph{``c''} and \emph{``B''} in Figure~\ref{fig:img}-{\it right} 
for the SF-QM and GS-QM methods, respectively.  
For the PCC method, the vertical axis corresponds to residuals between the 
spatial profiles, whereas for the QM method it corresponds to standard 
deviation.  
Units in the vertical axis have been normalized and shifted
for easy comparison.  
The vertical blue dashed lines mark the best-fit expansion for
each method.  
\label{fig:prof}
}
\end{figure}

An inspection of the July 2015 and May 2017 images reveals the nova 
expansion.  
To quantify the current expansion of J210204 and to investigate possible 
anisotropies, the images from these two epochs were compared using
different methods.  
As a preliminary step, the images were accurately registered using
30 reference stars in the FoV, the pixel size of the 2017 images  
(0\farcs211 pix$^{-1}$) was scaled to that of the 2015 images 
(0\farcs189 pix$^{-1}$), and the surface brightness normalized 
to the nebular peak value.

\begin{table}
\caption{J210204 expansion measurements} 
	\centering
	\label{tab:example_table}
	\begin{tabular}{ccccc}
		\hline
		\hline
          Feature & Emission & Angular & Expansion & Magnification   \\
		 Label     & Line         & Expansion & Rate & $f$  \\
		               &                 &  ($^{\prime\prime}$)     & ($^{\prime\prime}$ yr$^{-1}$) &  \\    
	\hline
	\multicolumn{5}{c}{PCC $\equiv$ Profile Cross-Correlation } \\
	\hline
 1	& [N {\sc ii}] &	0.142	    &	0.0753  &	0.0043 \\
 2  & [N {\sc ii}] &	0.203	    &	0.1075  &	0.0066 \\
 3 	&  [N {\sc ii}] &	0.150	    &	0.0794  &	0.0067 \\
 4 	&  [N {\sc ii}] &	0.116	    &	0.0613  &	0.0047 \\
 5 	&  [N {\sc ii}] &	0.068	    &	0.0361  &	0.0051 \\
 6 	&  [N {\sc ii}] &	0.156	    &	0.0824  &	0.0079 \\
 7 	&  [N {\sc ii}] &	0.161	    &	0.0854  &	0.0065 \\
 8 	&  [N {\sc ii}] &	0.173	    &	0.0914  &	0.0057 \\
 9 	&  [N {\sc ii}] &	0.213	    &	0.1125  &	0.0061 \\
10 &  [N {\sc ii}] &	0.133	    &	0.0703  &	0.0048 \\
11 	&  [N {\sc ii}] & 0.221	    &	0.1166  &	0.0083 \\
12 &  [O {\sc iii}] &	0.127	    &	0.0673  &	0.0026 \\
     \hline
     \multicolumn{5}{c}{SF-QM $\equiv$ Single Feature Quantified Magnification } \\
	\hline
a 	&  [N {\sc ii}] &	0.148	    &	0.0784  &	0.0044 \\	
 b 	&  [N {\sc ii}] &	0.110	    &	0.0583  &	0.0036 \\
 c 	&  [N {\sc ii}] &	0.136	    &	0.0723  &	0.0061 \\
 d 	&  [N {\sc ii}] &	0.093	    &	0.0492  &	0.0038 \\
 e 	&  [N {\sc ii}] &	0.079	    &	0.0422  &	0.0059 \\
 f 	&  [N {\sc ii}] &	0.134	    &	0.0713  &	0.0072 \\
 g 	&  [N {\sc ii}] &	0.134	    &	0.0713  &	0.0056 \\
 h 	&  [N {\sc ii}] &	0.169	    &	0.0894  &	0.0057 \\
 i 	&  [N {\sc ii}] &	0.212	    &	0.1125  &	0.0061 \\
 j 	&  [N {\sc ii}] &	0.144	    &	0.0764  &	0.0053 \\
 k 	&  [N {\sc ii}] &	0.115	    &	0.0613  &	0.0044 \\	
 \hline
 \multicolumn{5}{c}{GS-QM $\equiv$ Global Structure Quantified Magnification } \\
	\hline
 A 	&  [N {\sc ii}] &	0.136	    &	0.0723  &	0.0045 \\
 B 	&  [N {\sc ii}] &	0.114	    &	0.0603  &	0.0052 \\
 C &  [N {\sc ii}] &	0.096	    &	0.0512  &	0.0059 \\
 D &  [N {\sc ii}] &	0.115	    &	0.0613  &	0.0068 \\
 E 	&  [N {\sc ii}] &	0.152	    &	0.0804  &	0.0052 \\
 G &  [O {\sc iii}] &	0.108	    &	0.0573  &	0.0021 \\
 H &  [O {\sc iii}] &	0.098	    &	0.0522  &	0.0021 \\
  \hline
	\end{tabular}
	\label{tab:exp}
\end{table}

The first method consisted in the cross-correlation of radial profiles
extracted along directions chosen to include prominent knots and
filaments.
We will refer to it as the Profile Cross-Correlation (PCC) method.
A set of twelve features were considered, as labeled in 
Figure~\ref{fig:img}-{\it left}.  
The radial profiles extracted from the 2015 images were then
shifted and its difference with respect to the profiles from
the 2017 images computed to derive the shift that minimizes
this difference, as illustrated in Figure~\ref{fig:prof}.  
The expansion rate for the aforementioned features are listed in
Table~\ref{tab:exp}, where features 1-11 correspond to the 
[N~{\sc ii}] image and feature 12 to the [O~{\sc iii}] image.

The second method consisted in the minimization of residuals
between images of different epochs.  
In this method, that will be referred to as the Quantified Magnification
(QM) method, the 2015 images are magnified by a factor 1+\emph{f} and
then subtracted from the corresponding 2017 images.
The best magnification factor \emph{f} is derived by computing the
statistics in the difference image in boxes around prominent features
and minimizing the dispersion (Fig.~\ref{fig:prof}), i.e., looking
for the difference image with the smallest noise.  
This procedure is methodologically similar to that
applied by \citet{Szyszka_etal2011} using the IDL
routine MPFIT \citep{Markwardt2009} to the planetary
nebula NGC\,6302.  
Eleven small boxes around the single features (SF-QM method) labeled with
lowercase letters in Figure~\ref{fig:img}-{\it right}, and eight large
boxes around global structures (GS-QM method) labeled with uppercase
letters in the same figure were considered.  
The magnification factors for these boxes are listed in
Table~\ref{tab:exp}, where \emph{a}-\emph{k} correspond to
single features in [N~{\sc ii}]\footnote{
For a fair comparison between the PCC and QM methods, the 11 features 
selected for the PCC method correspond to the 11 features selected for 
the SF-QM method.}, \emph{A}-\emph{F} to global
structures in [N~{\sc ii}], and \emph{G}-\emph{H} to global 
structures in [O~{\sc iii}].

Table~\ref{tab:exp} compiles the magnification factors \emph{f} and
expansion rates derived for different features of J210204 using the
PCC and QM methods.  
The averaged magnification factors for the [N~{\sc ii}] ring are 
0.0061$\pm$0.0013 for the PCC method, 
0.0053$\pm$0.0011 for the SF-QM method, and 
0.0057$\pm$0.0009 for the GS-QM method, 
i.e.\ they agree within the 1-$\sigma$ standard deviation.  
The comparison between these methods is illustrated in Figure~\ref{fig:prof} 
for the particular case of knot \emph{3}$\equiv$\emph{c}$\equiv$\emph{B}, 
for which the three methods provides very similar values of the expansion 
rate.  
This is the general rule, but features \emph{2}$\equiv$\emph{b} 
and \emph{11}$\equiv$\emph{k} exhibit very discrepant values of 
the expansion rates, which seem to be associated with noticeable 
morphological changes in the features between the two epochs.

\begin{figure}
\centering
\includegraphics[bb=60 80 760 710,width=0.515\textwidth]{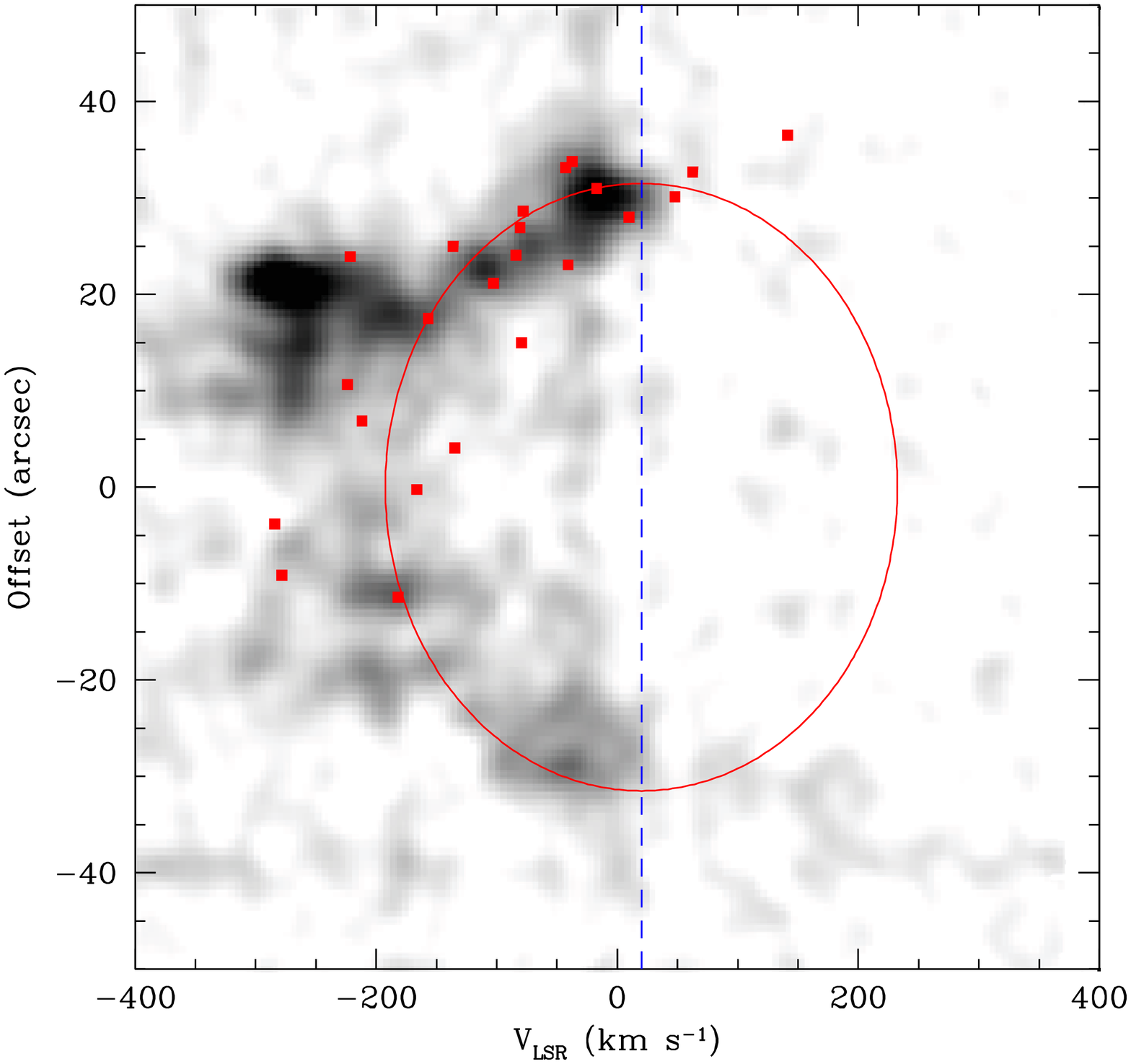}
\caption{
OAN-SPM 2.1m MES gray-scale position-velocity (PV) map over-plotted
with GTC OSIRIS data points (red squares).  
The blue dotted line marks the LSR radial velocity of J210204.  
The solid red ellipse is the best-fit model to the expansion
of the ring of J210204.  
  \label{fig:kin}}
\end{figure}

The averaged magnification factor is 0.0057$\pm$0.0012 for the ring 
and 0.0023$\pm$0.0003 for the bow-shock.  
The small dispersion of the magnification factor of the ring, 
$\approx$20\% the value of the magnification factor, implies 
that the ring expands mostly coherently.  
It also implies that there are no significant anisotropies.  
Otherwise, there is a noticeable difference for the expansion rate 
derived for the ring and the bow-shock.  
At the tip of the ring, the magnification factor implies an 
angular expansion rate 0\farcs100$\pm$0\farcs021~yr$^{-1}$,  
while it is 0\farcs059$\pm$0\farcs008~yr$^{-1}$ for the bow-shock.

\section{Nova Expansion Velocity and Distance} \label{sec:kinematics}

  The GTC OSIRIS and OAN-SPM MES data used in this work can be used
  to build a spatio-kinematics model of the ring of J210204 aiming
  at deriving an averaged expansion velocity.  
The MES 6\farcm5-long slit, placed across the main ring of
J210204 along of position angle of --13$^\circ$, results in
the position-velocity (PV) map shown in Figure~\ref{fig:kin}.  
The OSIRIS slits A and C presented by \citet{Guerrero_etal2018} contain
information on the region of the ring probed by the MES slit, thus the 
velocity of discrete knots along these OSIRIS slits has been measured,
corrected to the local standard of rest (LSR) system, and over-plotted
on the MES PV map in Figure~\ref{fig:kin}.
We note that the MES data have higher resolution than the
OSIRIS data, but the latter have much higher sensitivity.

  The kinematics derived from these data confirm the high-velocity of the
  knots.
  Compared to the systemic velocity of the nova shell, which is here
  assumed to be that of the central star, $V_\mathrm{LSR}=+17.5$ km~s$^{-1}$ 
  \citep{Guerrero_etal2018}, the knots in the ring of J210204 are mostly
  distributed along an arc with radial systemic velocities up to $-200$
  km~s$^{-1}$.
  Although the MES PV map reveals features with notably higher expansion
  velocities, up to $-300$ km~s$^{-1}$, the general kinematics and morphology
  of the [N~{\sc ii}] ring are coherent, implying ordered motion patterns.
  Following other spatio-kinematics models of nova shells, two basic
  geometrical models can be adopted to describe the observed kinematics
  and morphology of this structure:
  a homologous expanding prolate ellipsoid \citep[as in DQ\,Her,][]{VOR2007}
  or a flat ring expanding at constant speed \citep[as the equatorial rings of
  the bipolar nova shells FH\,Ser and HR\,Del,][]{GO2000,HO2003}.  
  In the former case, a noticeable line tilt at the tip of the major axis
  would be expected \citep[as in the Ring Nebula,][]{Guerrero_etal1997},
  thus we have adopted a model consisting of an expanding flat circular
  ring.
  Note that an oblate ellipsoid would result in a very similar
  velocity field and morphology.

  The ring inclination angle with respect to the line of sight has been
  adopted to be 53$^\circ\pm$8$^\circ$, as derived from the observed ring 
  aspect ratio, assuming it is the elliptical projection onto the sky of
  a circular ring.
	
The data points and PV maps in Figure~\ref{fig:kin} have then
been fitted using a least-square minimization method.
The expansion velocity is derived to be 285$\pm$30 km~s$^{-1}$, where the
error-bar already accounts for the uncertainty in the system inclination.
The expansion velocity of the ring, in conjunction with its
angular expansion, result in a distance of 0.60$\pm$0.13 kpc.
GAIA DR-2 provides a parallax for the central star of J210204 (source
ID \#2163877198882886656) of $\pi$=1.348$\pm$0.037 milli-arcsec,
i.e.\ a distance of 0.74$\pm$0.03 kpc, which is
  very similar to the value derived in this work.

\section{Long-term Nova Expansion and Age} \label{sec:yesterday_expansion}

\begin{figure}
\centering
\includegraphics[bb=30 115 660 582,width=0.975\columnwidth]{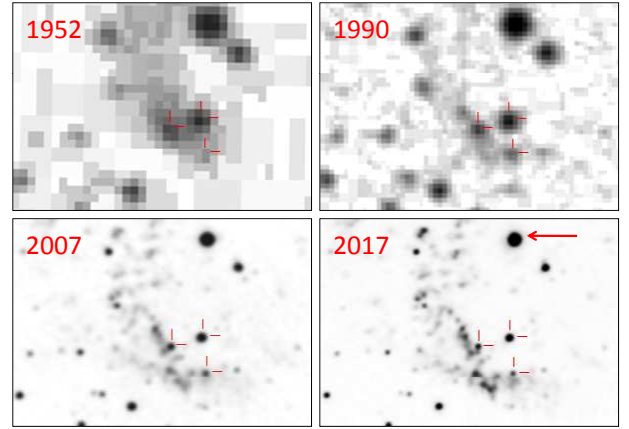}
\vspace*{-0.35cm}
\caption{
Gray-scale image of the Southernmost section of nova J210204 as seen 
in 1952 (POSS-I-E), 1990 (POSS-II-F), 2007 (INT IPHAS), and 2017 (NOT).
North is up, east to the left.  
The nova expansion can be compared with fiducial background stars
marked by red lines.
The central star of J210204 is marked by a red arrow.
\label{fig:fig4}
}
\end{figure}

\begin{figure*}
\centering
\includegraphics[width=1.5\columnwidth]{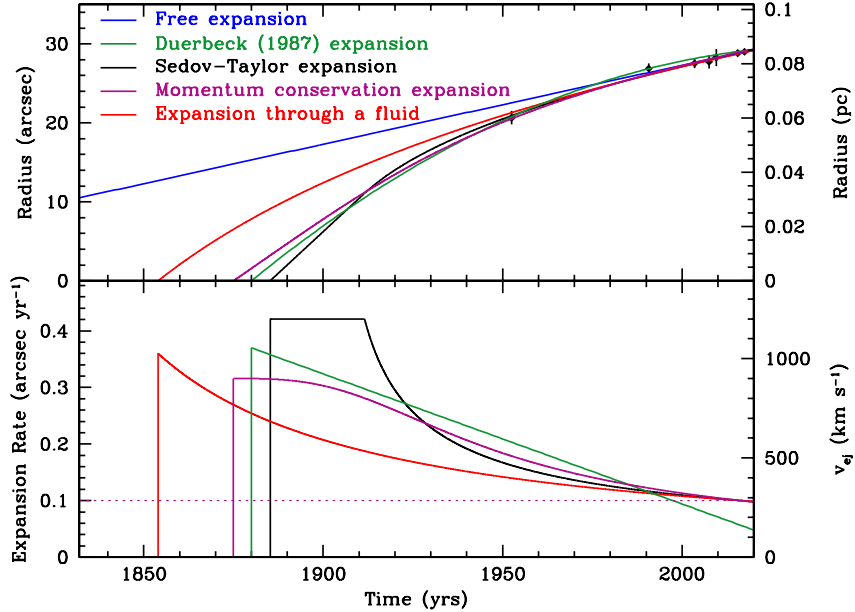}
\caption{Size of the semi-major axis of J210204 at different epochs
  overlaid by the best-fit models of the nova expansion (top) and
  expansion velocity of the different models (bottom).  Each data
  point in the top panel correspond to the average of the semi-major
  axis measurements obtained from a number, typically 3 to 4, of
  features in the ring of J210204 for each epoch.  The error bars in
  the semi-major axis of the top panel correspond to the dispersion of
  the individual values obtained for each epoch.
\label{fig:fig5}
}
\end{figure*}

The long-term expansion of J210204 is illustrated for the ring Southernmost 
tip in Figure~\ref{fig:fig4} using 1952 and 1990 DSS, 2007 INT, and 2017 NOT 
images.  
The measurement of the expansion rate, however, is hampered by the 
different plate scale and spatial resolution of these images.  
To overcome this problem, radial profiles along selected directions
including isolated knots in the ring were extracted and the distances
of these knots to the central star were then determined.  
Adopting for the ring the same geometrical model as in \S4, i.e.\ the ring 
is a flat expanding circular ring tilted to the line of sight by 53$^\circ$, 
the position angle on the sky of each of these knots and their radial 
distances can be used to derive the semi-major axis of the ring at each 
epoch.  
The time evolution of the semi-major axis at different epochs 
is plotted in the top-panel of Figure~\ref{fig:fig5}.

A free expansion ($v_\mathrm{exp}=cte$) of J210204 would imply
the linear increase of radius with time shown by the blue line
in Figure~\ref{fig:fig5}-top, for a nova event time circa 1730.  
The radius of J210204 in 1952, below this prediction,
indicates that the expansion of the ejecta is not free,
but it has slowed down with time.  
Furthermore, a free expansion would imply an expansion velocity of 
285~km~s$^{-1}$, which is lower than the typical initial expansion 
velocity of novae.  
Different expansions laws are investigated next to determine
the nova age from the observed angular expansion with time.

\citet{Duerbeck1987} assumed a quadratic law to describe a nova expansion: 
\begin{equation} \hspace{2cm}
r = v_0\times(t-t_0) + a\times(t-t_0)^2
\end{equation}
The best fit to the data using this expansion law (green in
Fig.~\ref{fig:fig5}) suggests that the nova event took place 
circa 1880 with initial expansion velocity 1050 km~s$^{-1}$.
Whereas an expansion law with a constant deceleration does not have a 
physical support, we note that the fit underestimates the current 
expansion rate of J210204 (lower panel of Fig.~\ref{fig:fig5}).

In a supernova there is an initial phase of free expansion followed by a
deceleration phase, the Sedov-Taylor phase, which starts when the swept
mass is greater than the ejected mass.  
The expansion is described by the following expressions
\citep{Sedov1959,Taylor1950}:
\begin{equation} \hspace{2cm}
r \propto (t - t_0)^{-0.6} ; \;\;\; v \propto (t - t_0)^{0.4}, 
\end{equation}
where an initial expansion velocity of 1200 km~s$^{-1}$ 
has been assumed.  
Under this assumption, the phase of free expansion lasts $\sim$25 yr 
(black lines in Figure~\ref{fig:fig5}), whereas in supernovae this phase 
may last up to 1000 yr.  
The nova event is estimated to have happened circa 1885.

In late evolutionary phases of supernovae the ejecta
cools down and expands against the ISM due to its own
inertia.
If we assume J210204 already entered this so-called momentum
conservation or snow-plow phase, its initial mass $m_0$ and
initial and present expansion velocities $v_0$ and $v$ are
related to the medium density $\rho_0$ as:
\begin{equation} \hspace{2.3cm}
m_0 v_0 = (m_0 + \dfrac{4\pi}{3} \rho_0 r^3 ) v  
\end{equation}
resulting in the purple line in Figure~\ref{fig:fig5},
which implies the nova event took place in 1875 and 
had an initial expansion velocity of 900 km~s$^{-1}$.

Finally, if J210204 were expanding as discrete bullets moving through 
the ISM \citep{Williams2013}, its motion could be described by the drag 
equation:
\begin{equation} \hspace{3.4cm}
a \propto -v^2
\end{equation}
The fit of this equation to the data points is shown in red
in Figure~\ref{fig:fig5}), for a nova event circa 1850.

\subsection{Historical search}

Adopting either a typical absolute magnitude $M_\mathrm{v}$=--7 mag
for a CNe \citep{Warner1987} at the distance of 0.60 kpc, or an
average brightening of $\sim$11 mag \citep{BE2008} from its present
$m_\mathrm{v}$=15.8 mag, J210204 would have been at
$m_\mathrm{v}$=1.9--4.8 mag at peak brightness.  Such a nova should
have been detectable to the naked eye.  Although none of the expansion
laws considered above produce a perfect match to the variation of
radius with time and to the present expansion rate of J210204, a
consistent time for the nova event between 1850 and 1890 is achieved
by different models accounting for the deceleration of the ejecta.

At that time, most of the brighter novae continued to be found 
by amateurs, although the use of wide-field photography was 
introduced and extended in professional observatories in the 
last two decades of the 19th century.
We searched evidence for diffuse emission from the nova shell in the
Heidelberg Digitized Astronomical Plates taken at least half a century
after the nova event.
No detection can be claimed, but it must be noted that most of
these images were obtained through blue filters or using blue
sensitive plates, which are not suited for the prevalent 
H$\alpha$ and [N~{\sc ii}] (red) emission lines of J210204.

We have examined the most complete list of Galactic novae 
provided at https://projectpluto.com/galnovae/galnovae.htm, 
which includes a small sample of objects discovered before 1900.  
Apparently there is no record of this event.
Only eight bright novae were visible to the naked eye in the 19th century 
\citep[with T\,CrB and Q\,Cyg discovered in 1866 and 1876 at 2.0 and 3.0 
mag, respectively, among the brightest ones;][]{Warner2006}, compared to 
the 30 bright novae visible to the naked eye in the 20th century.
  Assuming similar nova rate for the 19th and 20th centuries, the figures
  above imply that $\sim$75\% of novae visible to the naked eye were
  missed in the 19th century.
It looks like J210204 is one of those novae missed at their time.

\section{Discussion}

The analysis of the long-term expansion of J210204 presented in the 
previous section certainly discards its free expansion.  
On the other hand, all expansion laws accounting for the deceleration
of the ejecta imply consistent times for the nova event between 1850
and 1890, and initial expansion velocities in the range 1000--1500
km~s$^{-1}$ typical of novae.  
Since then, the expansion of the main component of the nova ejecta, the 
[N~{\sc ii}]-bright arc, has decelerated down to the current expansion
speed of 285 km~s$^{-1}$.  
The situation could be more extreme for the bow-shock, 
whose larger distance from the central star and lower 
angular expansion would imply a stronger deceleration.  
Its chemical composition, mostly consistent with that of the 
ISM \citep{Guerrero_etal2018}, indeed suggests it has swept 
up a significant volume of the surrounding medium.

The expansion and deceleration of nova shells has received little 
attention in the past.  
It is a difficult task that requires the availability of 
high-quality multi-epoch images with a suitably long time 
baseline.  
For instance, \citet{Shara_etal2012a} could only determine an upper 
limit of $\leq$0\farcs17~yr$^{-1}$ for the expansion of the Z\,Cam
shell using images obtained in 2007 and 2010, i.e.\ a time baseline
of 3 yrs.  
\citet{Duerbeck1987} compared photographic plates obtained around 1950 with 
more recent (mid 80's) CCD images of four novae (DQ\,Her, GK\,Per, V476\,Cyg, 
and V603\,Aql) to conclude that they decelerate with a mean half-times of 75 
yrs, i.e.\ the expansion velocity of a nova drops to half its value every 75 
yrs.  
  Further evidence for shell deceleration is presented by \citet{DD2000}.
This expansion behavior is consistent with \citet{Oort1946} theoretical
estimate based on a simple model of the nova shell interaction with the
surrounding ISM.

These results, however, have been disputed for the nova shells
GK\,Per and DQ\,Her.  
GK\,Per nova shell has multiple knots expanding at an angular velocity 
0\farcs3-0\farcs5~yr$^{-1}$ that has remained unchanged since 
their ejection about a century ago \citep{AK2005,Liimets_etal2012}.  
The interaction of these knots with the wind emanating from GK\,Per has been 
claimed responsible for their kinematic behaviour \citep{Shara_etal2012b}.  
This is also the case for DQ\,Her nova shell, where the presence of
ablated flows associated with clumps, with notable tails extending
outwards with increasing radial velocities, most likely implies the
action of a stellar wind that speeds up the material in the nova
shell \citep{VOR2007}.  
More and higher quality images of nova shells spanning over 
long time baselines are most required to investigate the 
interaction of nova shells with their ambient medium.

  Otherwise, many high quality
studies of the kinematics of nova shells disclose overall expansion
patterns, 3-D structures, and dynamical interactions
as well as undeniable bipolar structures 
\citep[e.g.][]{dValle_etal1997,GO2000,HO2003,VOR2007,MD2009,Woudt2009,Shara_etal2012b,Shara_etal2017}.

It is not clear, however, what are the key mechanisms behind the
ejection of bipolar structures in nova shells \citep[for example,
the shaping is attributed to the binary companion or to rotation
of the WD envelope;][]{Lloyd1997,Porter1998}, but it is accepted
that it produces an interaction of the fast ejecta with slow-moving
or stationary material \citep[see,  e.g.,][]{Chomiuk2014}.
Although there are not many detailed numerical works in the literature 
addressing these interactions, the consequences of the bipolar
ejections on the observational properties of nova shells have been
discussed thoroughly \citep{Shore2013}.
This interaction piles up material along the equatorial plane, creating
dense ring-like structures around the binary system \citep{Chochol1997}.
Along with these lines, the morphology of the bright [N~{\sc ii}]
arc of J210204 (or AT\,Cnc) can be matched by the simple models of
nova shells presented by \citet{GO1999} for cases of bright rings close 
to the equatorial plane and faint polar caps.

Our combined analysis of the angular expansion and radial velocity of
the [N~{\sc ii}]-bright arc of J210204 indicate that the overall
expansion of the different knots is not completely coherent, as shown
by their varying magnification factors and by the complex kinematics.
Nonetheless, these variations are not as large as suggested by the
intricate morphology of this [N~{\sc ii}]-bright arc.  The angular
expansion and kinematics of the nova ejecta provide a value for the
distance to J210204 which is consistent with that derived from GAIA
DR-2, although not exactly the same. The difference may be due to
dynamical effects.

We suggest that, similarly to other astrophysical systems, the interaction
of the current stellar wind of the central star of J210204 with the ring
structure causes hydrodynamical instabilities, most likely of Rayleigh-Taylor
nature.
This could break the ring-structure into clumps that would further develop
cometary tails \citep[see][]{Fang2014,Harvey_etal2016}.
Detailed hydro-dynamical modeling to interpret the morphological
evolution of the knots and to investigate the dynamical evolution
of the nova ejecta will be presented in a future paper (Toal\'a et al. in prep.).

To summarize, J210204 is a CN shell resulting from a nova event that
took place about 130-170 yrs ago, although it was unnoticed at that
time.  The nova has experienced a notable deceleration, most likely
due to its occurrence inside a high density medium as suggested by the
complex H$\alpha$ emission around it \citep{Guerrero_etal2018}.  As
many other ``young'' novae, the CV at its center is in a nova-like
(NL) stage \citep{Guerrero_etal2018} with the WD still experiencing
accretion at a high rate \citep{Collazzi_etal2009}.

\section*{Acknowledgments} 

E.S.\ acknowledges support from Universidad de Guadalajara and
CONACyT.  M.A.G.\ acknowledges support of the grant AYA2014- 57280-P,
cofunded with FEDER funds.
G.R.-L.\ acknowledges support from Fundaci\'on Marcos Moshinsky,
CONACyT, and PRODEP (Mexico).
L.S.\ acknowledges support from PAPIIT grant IA-101316 (Mexico).
J.A.T.\ and M.A.G.\ are funded by UNAM DGAPA PAPIIT project IA100318.
We thank Martin Henze for his advise on historical records of novae
and photographic plates.

This article is based upon observations obtained with ALFOSC, which is 
provided by the Instituto de Astrof\'\i sica de Andaluc\'\i a (IAA) 
under a joint agreement with the University of Copenhagen and NOTSA, 
and with OSIRIS at the Gran Telescopio Canarias (GTC), both installed 
in the Spanish Observatorio del Roque de los Muchachos of the Instituto 
de Astrof\'\i sica de Canarias, in the island of La Palma, Spain.  
Observations carried out at the Observatorio Astron\'omico Nacional 
on the Sierra San Pedro M\'artir (OAN-SPM), Baja California, M\'exico, 
were also used.

This paper also makes use of data obtained as part of the INT Photometric 
H$\alpha$ Survey of the Northern Galactic Plane (IPHAS: http://www.iphas.org) 
carried out at the Isaac Newton Telescope (INT). 
The INT is operated on the island of La Palma by the Isaac Newton Group 
in the Spanish Observatorio del Roque de los Muchachos of the Instituto 
de Astrof\'\i sica de Canarias. 
All IPHAS data are processed by the Cambridge Astronomical Survey Unit 
at the Institute of Astronomy in Cambridge. 
The band-merged DR2 catalog was assembled at the Centre for Astrophysics 
Research, University of Hertfordshire, supported by STFC grant ST/J001333/1.
The Digitized Sky Surveys were produced at the Space Telescope Science
Institute (STScI) under US Government grant NAGW-2166.
The images of these surveys are based on photographic data obtained using
the Oschin Schmidt Telescope on Palomar Mountain and the UK Schmidt
Telescope.
The plates were processed into the present compressed digital form with
the permission of these institutions.
The National Geographic Society - Palomar Observatory Sky Atlas (POSS-I)
was made by the California Institute of Technology with grants from the
National Geographic Society.
The second Palomar Observatory Sky Atlas (POSS-II) was made by the
California Institute of Technology with funds from the National
Science Foundation, the National Geographic Society, the Sloan
Foundation, the Samuel Oschin Foundation and the Eastman Kodak
Corporation.
The Oschin Schmidt Telescope is operated by the California
Institute of Technology and Palomar Observatory.
The UK Schmidt Telescope was operated by the Royal Observatory,
Edinburgh, with funding from the UK Science and Engineering
Research Council (later the UK Particle Physics and Astronomy
Research Council), until 1988 June, and thereafter by the
Anglo-Australian Observatory. Supplemental funding for sky-survey
work at the STScI is provided by the European Southern Observatory.
This work made use of the HDAP which was produced at Landessternwarte 
Heidelberg-K\"onigstuhl under grant No.\ 00.071.2005 of the 
Klaus-Tschira-Foundation.

%%%%%%%%%%%%%%%%%%%%%%%%%%%%%%%%%%%%%%%%%%%%%%%%%%



%\bibliographystyle{mnras}


%%%%%%%%%%%%%%%%%%%%%%%%%%%%%%%%%%%%%%%%%%%%%%%%%%

%%%%%%%%%%%%%%%%% APPENDICES %%%%%%%%%%%%%%%%%%%%%

%\appendix



%%%%%%%%%%%%%%%%%%%%%%%%%%%%%%%%%%%%%%%%%%%%%%%%%%

% Don't change these lines
\bsp	% typesetting comment
\label{lastpage}
\end{document}